# Whole Slide Image to DICOM Conversion as Event-Driven Cloud Infrastructure

**Authors:**

David Brundage[1,2,*], Jacob Rosenthal[1,2], Ryan Carelli[1,2], Sophie Rand[1,2], Renato Umeton[1,2,3,4], Massimo Loda[1], Luigi Marchionni[1]

**Affiliations:**

1. Weill Cornell Medicine, Department of Pathology and Laboratory Medicine
2. Dana-Farber Cancer Institute, Department of Informatics and Analytics
3. Harvard T.H. Chan School of Public Health, Department of Biostatistics
4. Massachusetts Institute of Technology

* corresponding author

**Abstract**

The Digital Imaging and Communication in Medicine (DICOM) specification is increasingly being adopted in digital pathology to promote data standardization and interoperability. Efficient conversion of proprietary file formats into the DICOM standard format is a key requirement for institutional adoption of DICOM, necessary to ensure compatibility with existing scanners, microscopes, and data archives. Here, we present a cloud computing architecture for DICOM conversion, leveraging an event-driven microservices framework hosted in a serverless computing environment in Google Cloud to enable efficient DICOM conversion at scales ranging from individual images to institutional-scale datasets. In our experiments, employing a microservices-based approach substantially reduced runtime to process a batch of images relative to parallel and serial processing. This work demonstrates the importance of designing scalable systems for enabling enterprise-level adoption of digital pathology workflows, and provides a blueprint for using a microservice architecture to enable efficient DICOM conversion.

# Introduction

Digital Imaging and Communication in Medicine (DICOM) has been the standard communication and file format for medical imaging data due to its adoption in radiology for multiple modalities[1]. The DICOM standard has enabled rapid expansion in the development of interoperable systems and artificial intelligence in the field of radiology. While there have been similar efforts to adopt the DICOM specification in the context of digital pathology to enable storing whole slide images (WSI) [2], including implementation of requirements specific to digital pathology, adoption of the DICOM standard in pathology has lagged.

This lag in adoption has also been seen in industry-standard slide scanners exporting images in proprietary file formats. These proprietary file formats can hinder the adoption of digital pathology by creating vendor lock-in and reducing technical stack choice to only vendor-supported solutions. Even as some commercially available scanners come to market that support DICOM natively, such as the Aperio GT450 [3], adoption has continued to crawl because laboratories' cost to update their scanning infrastructure. Another challenge to adoption is that while DICOM may be undergoing slow but sure increases in adoption in pathology, proprietary file formats commonly used today must continue to be supported for integrating the vast number of slides that have already been digitized, allowing support for legacy devices. Both of these costs - of conversion from legacy scanners to instruments that support DICOM; and cost of conversion of legacy file formats to the DICOM scanner represent large barriers to adoption. Therefore, infrastructure for converting WSIs from proprietary file formats natively output from scanners and microscopes into the DICOM standard format is a critical piece of moving the field towards full adoption of DICOM.

The process of converting the legacy images to DICOM is computationally expensive, primarily due to the large size of high-resolution gigapixel WSIs. These large files often cannot be loaded into memory all at once, requiring purpose-built tools designed to handle images at such scale efficiently. As a result, several conversion applications have been developed, such as the Python-based implementations from Hermann et al. and Gu et al. [4, 5] as well as C++ converters based on OpenSlide library[6] from Orthanc [7] and Google [8].

Existing approaches to DICOM implementation for digital pathology have focused on developing and performing conversion applications running in a local environment. However, these implementations would not scale at the enterprise level in a network consisting of 11 hospitals and serving a population of more than 10 million patients, these existing workflows are not sufficiently scalable. Here, we present a cloud-based infrastructure to support the conversion of WSI to DICOM in an enterprise-scale production environment. Our pipeline consists of a microservice framework (Figure 1) to handle each component of the conversion process in a secure, scalable, and containerized format. The primary components of the pipeline consist of cloud storage for unprocessed raw data, a notification service to trigger data creation events, a publish-subscribe (pub/sub) service for event-driven message streaming, and autoscaling containerized web applications to run DICOM conversion for each image. This approach enables our application to be massively scalable, robust, and fault-tolerant while being modular and vendor agnostic, addressing an unmet need for enterprise-scale digital pathology infrastructure.

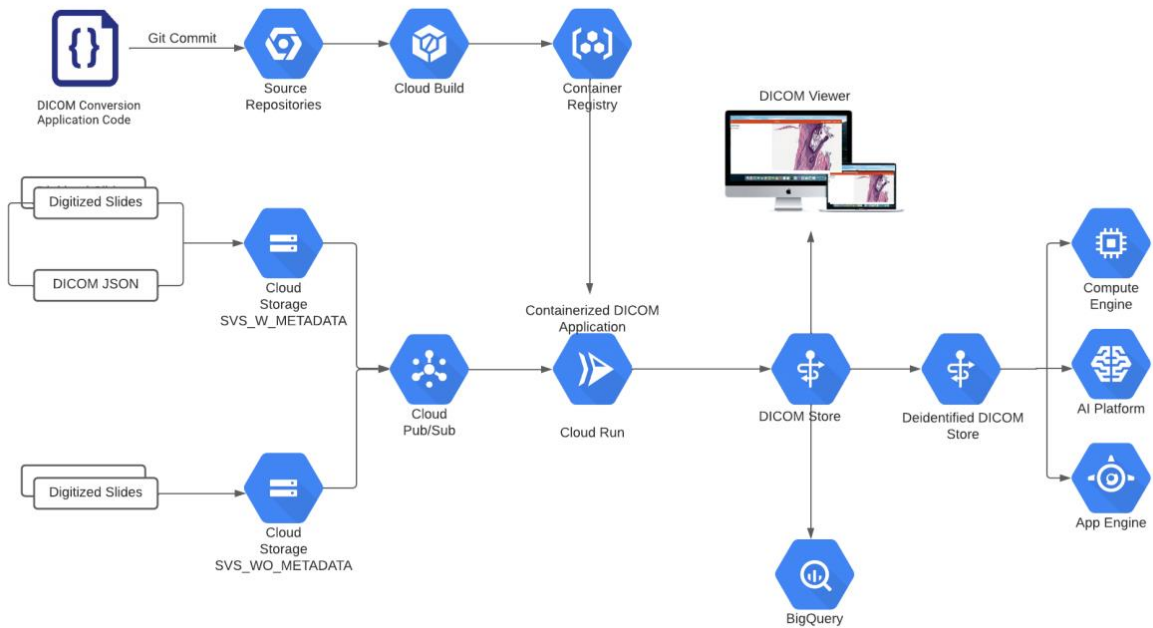

*Figure 1 Event-driven microservice architecture with destination and downstream application.*

## Methods

### Microservice Architecture

Microservice architecture is a design pattern that structures an application as a collection of stand-alone units. By decoupling independent components of the application, a microservice architecture can reduce cost, increase software release cycle speed, improve resilience, and enable visibility[9]. Because each service is self-contained and smaller in scope, microservice architectures enable the reliable delivery of large applications with the ability to be modified quickly and frequently by a small team with an evolving technology stack. By leveraging a microservice-style design, we can further improve operational and maintenance efficiency by embracing Infrastructure as Code [10].

Our WSI DICOM conversion application consists of 3 microservices: a publish-subscribe (pub/sub) messaging system, a cloud storage resource, and a DICOM conversion web application. All services log to a shared location which facilitates monitoring the entire application through a single dashboard. The cloud storage bucket acts as our publisher in the DICOM conversion infrastructure, sending a new event message to the WSI-DICOM conversion topic each time a new file is uploaded. The topic is a resource in the pub/sub architecture that receives messages from a publishing application. The messages capture event metadata such as date and time of image creation and a pointer to its location in cloud storage. This metadata is passed as the payload information in the body of a restful API request to the event target. The subscriber application runs a containerized DICOM conversion application. Our subscription utilizes a push architecture for message delivery, allowing it to deliver messages to an HTTPS endpoint for our subscriber application as they are received in real-time. The final application sends the converted image to the enterprise DICOM store. In the following sections, we detail each service individually.

**Cloud Storage and Ingestion Service**

In the DICOM conversion infrastructure application, the cloud storage bucket is the initial landing zone from on-premise data services (i.e., slide scanners). When a new file enters the landing zone in the bucket, our service generates an event notification for object creation and sends the notification to the pub/sub topic.

One benefit of cloud infrastructure is decoupling storage from computational resources. By storing the WSI in its vendor native format in cloud storage, the original files can be retrieved as often as needed in low latency, highly durable, and geo-redundant medium. The raw object store reduces costs by leveraging storage classes without having to sacrifice performance or data availability. Policies on the bucket move the WSI to cold storage based on lifecycle definitions and archival storage based on institutional requirements for retention.

**Publish-Subscribe Architecture and Messaging Service**

Publish-subscribe (pub/sub) is an architecture pattern for transmitting information between the microservices and is used to enable orchestration of the application. We leverage a topic-based paradigm for our pub/sub service. Services publish messages to a topic, and other services can receive that information as it becomes available by subscribing to the topic. The topic acts as a named resource that groups specific information received by publishing applications. The subscription connects the topic to a subscriber application that can receive and process the messages that are published to the topic. A dedicated microservice manages the pub/sub messaging component.

The pub/sub architecture is a good fit for our use case of deploying institutional-scale data ingestion and conversion application because it allows decoupling between data producers and consumers, providing a scalable solution for exchanging information in a distributed environment. Allowing subscriptions to consume data from a specific topic lets the data be scaled out horizontally to multiple users without modifying the ingestion point. New hospitals can be brought online to the enterprise digital pathology architecture by simply adding an ingestion point and directing it to the appropriate pub/sub topic, without requiring modifications to any of the other components.

In addition, this event-driven paradigm provides a platform to integrate with other applications, as new services can subscribe to preexisting topics. Scalability is a key requirement for enabling the next generation of computational image-based cancer research[11], and this pub/sub architecture can be naturally extended to support other use cases beyond image conversion. For example, the Whole Slide Image ingestion topic could also send the slide notification to a machine learning model for processing or an administrative workflow for quality assurance. By creating decoupled points for data ingestion, we can reduce the need to stand up additional infrastructure, increasing delivery speed and allowing additional transparency into the service.

**Containerized Conversion Web Application Service**

The subscription application is a Docker [12] containerized Flask [13] web application that receives and processes messages from the pub/sub event subscription. The web application consists of two components. The first component ingests the HTTPS request from the event notification and parses the WSI information, including the triggered storage bucket and the originating file. Once the Flask application receives the message, it downloads the file from the cloud storage bucket and processes it using the chosen DICOM file converter. We chose a file converter written in C++, Google WSI [8], due to the processing speed gained from using a compiled language[4]. Because this entire pipeline is modular, the file conversion application can be separated from the HTTPS ingestion allowing institutions to use their conversion tool of choice. After the WSI has been converted to DICOM, the Flask application acknowledges the request with an HTTPS 200 response to the pub/sub subscription, indicating that the request has succeeded, and removes the message from the queue. Once the cycle of message ingestion, image processing, and acknowledgment has been completed, the application can begin receiving new messages.

The containerized application is hosted in an autoscaling serverless computing environment in Google Cloud Run. Using a serverless backend to host the application effectively disconnects the computation from where it will run, allowing us to abstract away some of the management of the underlying hardware. The number of servers, capacity, and provisioning can be modulated to meet workload demands, enabling autoscaling of computational resources based on utilization from zero to N. Scaling to zero allows the application to avoid costs when not in use, while scaling to N enables the application to dynamically allocate increased computational resources to handle bursts in processing demand. Since the compute costs are metered, users only pay for the time and resources consumed, reducing the costs from idle compute [14]. Leveraging a managed service allows the digital pathology developers to focus on the application's business logic instead of configuring and deploying servers and virtual machines.

## Results

**Workflows for Image Conversion**

To test the performance of our implementation, we converted 50 Whole Slide Images of the prostate in SVS format from The Cancer Genome Atlas (TCGA)[15] to the DICOM standard using the Google WSI2DCM tool. Each image was processed through 3 conversion workflows, and the total processing time was recorded after processing 1, 10, 25, and 50 images (figure 2). A cloud virtual machine with 16 vCPU and 64 GB of memory was used for processing the serial and parallel workflows. The serial processing workflow was completed by submitting a batch of images to the virtual machine, and each image was converted to DICOM sequentially. Parallel processing of images was a batch submission, and then Python's multiprocessing library was used to create a pool. The images were submitted to the pool with a function to call each conversion as a subprocess. Our autoscaling microservice workflow started with uploading the images to a cloud storage bucket triggering an object creation event notification to our message topic. A subscription on the topic pushed each notification event to the conversion microservice and scaled container instances to meet demand. Each container instance was designed to handle one request and convert one file to DICOM.

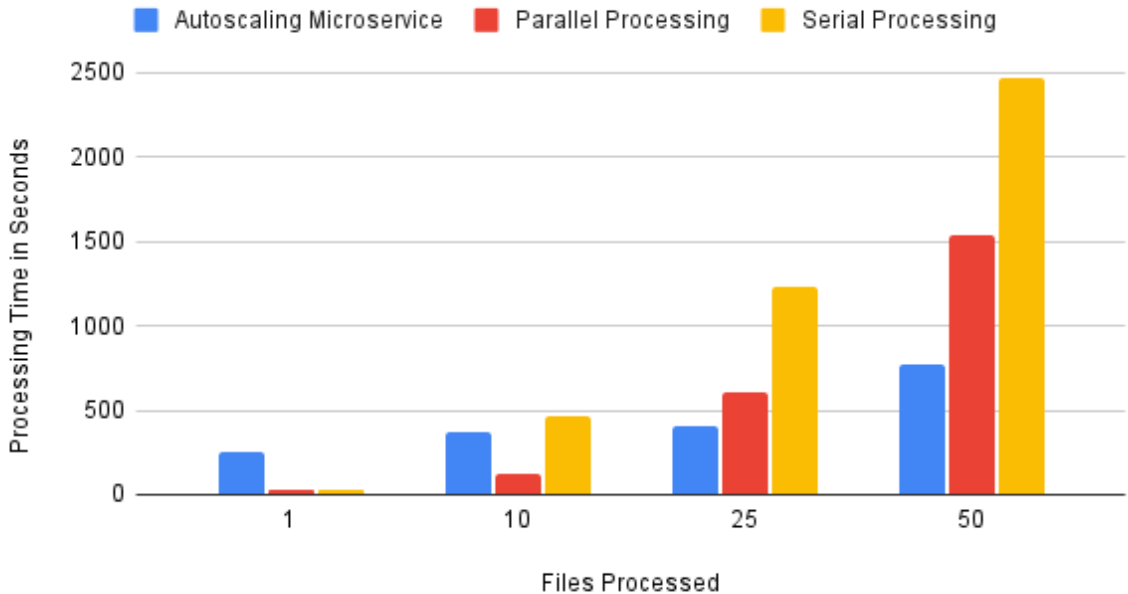

*Figure 2 Processing time in seconds to convert whole slide images to DICOM for autoscaling, parallel, and serial processing workflows.*

**Autoscaling and Limitations**

The architecture we propose does have limitations due to the cold start time as new containers are created. The cold start is a period when the container instance must be created to handle a request. This cold start time allows processing serially and in parallel to outperform the autoscaling infrastructure when processing one image, or only a few images, that cannot use the scalable container limit. Once the service has scaled to meet demands and the containers are hot, they can immediately serve requests, and our microservice-based architecture significantly outperforms both other workflows (Figure 2). In addition, the transparency and logging of our system allowed us to monitor in real-time as the service rapidly scaled up to meet the submitted requests, plateaued as all requests were serviced, and then slowly scaled the container instances back down to 0 (Figure 3). This feature allowed the service to handle the burst processing required for all 50 images while saving compute costs after processing had been completed.

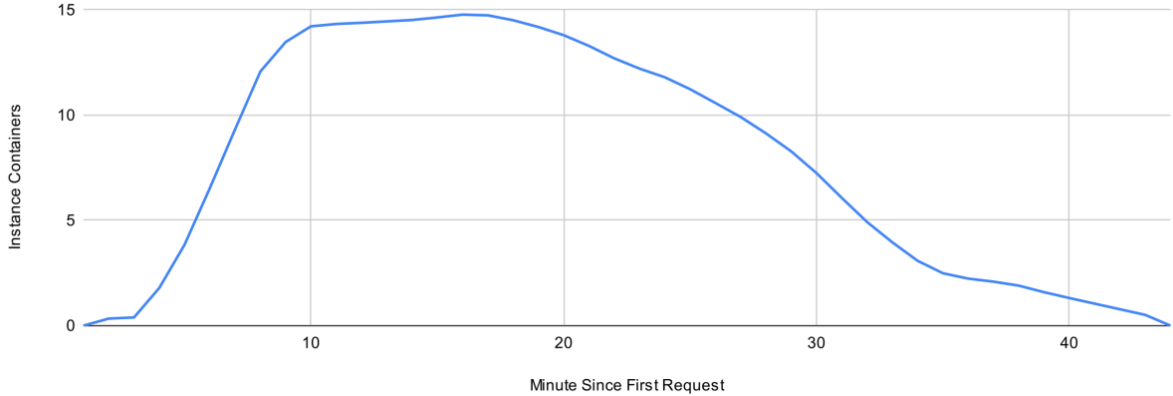

## Discussion

We have demonstrated an application for converting whole-slide images to the DICOM standard at an institutional scale. Implementing the DICOM standard and testing the performance of conversion applications has been the topic of multiple articles, but few have discussed an institutional approach to deploying these methods. We demonstrate a blueprint for developing scalable infrastructure to enable innovations in digital pathology at the system level.

Our implementation is not without limitations. First, this cloud architecture requires a base infrastructure to support security and compliance items for handling Personal Health Information (PHI). While all services used in the described architecture were covered under an approved business associate agreement (BAA) and HIPAA compliant, the individual institution must ensure that these services are configured appropriately.

Second, the serverless DICOM conversion containers have a cold start period before serving images for processing. This cold start period can be reduced by increasing the number of container instances that are always available. Though this is a potential solution, it will mean that the architecture will always have idle containers when no events are triggering the pipeline and increase computing costs. Therefore, a tradeoff on always available instances, the number of instances to scale to, and the number of concurrent requests a container should handle must be made.

Perhaps most importantly, adoption and implementation of an infrastructure to support whole slide image conversion requires a financial investment in both technical and personnel resources. Ensuring that the technicians supporting these pipelines are well versed in the technical infrastructure, as well as the clinical and operational implications are required for successful adoption of event drive microservices.

While there are still many challenges towards fully adopting the DICOM standard for digital pathology, this architecture demonstrates a solution enabling institutions to embrace the standardization of DICOM while still supporting legacy systems. Implementing autoscaling, event-driven, containerized services in pathology architecture enables an ecosystem that can be proactive in meeting the evolving needs and resource requirements to support digital services.

References

1. *DICOM Standard*. 2021; Available from: https://www.dicomstandard.org/current.
2. *DICOM Whole Slide Image (WSI)*. 2021.
3. Biosystems, L., *Aperio GT 450 - Atuomated, High Capacity Digital Pathology Slide Scanner*. 2021.
4. Gu, Q., et al., *Dicom_wsi: A python implementation for converting whole-slide images to digital imaging and Communications in Medicine compliant files.* Journal of Pathology Informatics, 2021. **12**(1): p. 21-21.
5. Herrmann, M., et al., *Implementing the DICOM standard for digital pathology.* Journal of Pathology Informatics, 2018. **9**(1): p. 37-37.
6. Goode, A., et al., *OpenSlide: A vendor-neutral software foundation for digital pathology.* Journal of Pathology Informatics, 2013. **4**(1): p. 27-27.
7. Jodogne, S., *The Orthanc Ecosystem for Medical Imaging.* Journal of Digital Imaging, 2018. **31**(3): p. 341-352.
8. Platform, G.C., *GoogleCloudPlatform/wsi-to-dicom-converter.* 2021.
9. Lauretis, L.D. *From Monolithic Architecture to Microservices Architecture*. in *2019 IEEE International Symposium on Software Reliability Engineering Workshops (ISSREW)*. 2019.
10. Kang, H., M. Le, and S. Tao. *Container and Microservice Driven Design for Cloud Infrastructure DevOps*. in *2016 IEEE International Conference on Cloud Engineering (IC2E)*. 2016.
11. Rosenthal, J., et al., *Building Tools for Machine Learning and Artificial Intelligence in Cancer Research: Best Practices and a Case Study with the PathML Toolkit for Computational Pathology.* Molecular Cancer Research, 2022. **20**(2): p. 202-206.
12. Docker, i., *Docker*. 2013.
13. Ronacher, A., *Flask Web Framework*. 2010.
14. Baldini, I., et al., *Serverless Computing: Current Trends and Open Problems*, in *Research Advances in Cloud Computing*, S. Chaudhary, G. Somani, and R. Buyya, Editors. 2017, Springer Singapore: Singapore. p. 1-20.
15. Cancer Genome Atlas Research, N., *The Molecular Taxonomy of Primary Prostate Cancer.* Cell, 2015. **163**(4): p. 1011-1025.